\UseRawInputEncoding
\documentclass[aps,prx,superscriptaddress,reprint]{revtex4-1}
\usepackage[usenames,dvipsnames]{xcolor}
\usepackage{lipsum}
\usepackage{amsmath}
\usepackage{amssymb}
\usepackage{graphicx}
\usepackage{url}
\usepackage{hyperref}
\usepackage{color,xcolor}
\usepackage{epstopdf}
\usepackage{float}
\usepackage{ulem}
\usepackage[percent]{overpic}
\usepackage{siunitx}
\usepackage{subdepth}

\usepackage{dcolumn}
\usepackage{bm}
\usepackage{times}
\usepackage[version=3]{mhchem} 

\usepackage{mathrsfs}
\usepackage{lipsum}
\usepackage{amsfonts}
\usepackage{array}
\usepackage{ulem}
\usepackage{lineno}
\usepackage[none]{hyphenat}

\hypersetup{colorlinks=true,citecolor=Red,linkcolor=Blue,urlcolor=Black}
\begin{document}
\title{Majorana zero modes in $Y$-shape interacting Kitaev wires}

\author{Bradraj Pandey}
\affiliation{Department of Physics and Astronomy, The University of 
Tennessee, Knoxville, Tennessee 37996, USA}
\affiliation{Materials Science and Technology Division, Oak Ridge National 
Laboratory, Oak Ridge, Tennessee 37831, USA}

\author{Nitin Kaushal}
\affiliation{Materials Science and Technology Division, Oak Ridge National 
Laboratory, Oak Ridge, Tennessee 37831, USA}

\author{Gonzalo Alvarez}
\affiliation{
Computational Sciences and Engineering Division, Oak Ridge, Tennessee 37831, USA}

\author{Elbio Dagotto}
\affiliation{Department of Physics and Astronomy, The University of 
Tennessee, Knoxville, Tennessee 37996, USA}
\affiliation{Materials Science and Technology Division, Oak Ridge National 
Laboratory, Oak Ridge, Tennessee 37831, USA}

\date{\today}
\begin{abstract}

Motivated by the recent experimental realization of minimal Kitaev chains using quantum dots, 
we investigate the  Majorana zero modes (MZM) in $Y$-shape Kitaev wires. 
We solve the associated  Kitaev models analytically at the sweet spot ($t_h=\Delta$) and derive the exact form of MZM
wave-functions in this geometry. {\ {The novelty of our result is the observation of 
{\it multi-site} MZMs located near the junction center, on the nearby edge sites of each leg. 
This result is important for potential braiding of Majoranas  and the performance of  $Y$-junctions
made from arrays of quantum dots.}} 
 Furthermore, we study the stability of local (single-site) and {\ {multi-site}} MZMs modes in the presence of Coulomb repulsion,
 using density matrix renormalization group theory. 
 Our local density-of-states calculation shows that these multi-site MZMs are as equally topologically protected as the 
single-site MZMs when in the presence of Coulomb repulsion {\ {or when away from the sweet-spot}}.
\end{abstract}
\maketitle

\noindent {\bf \\Introduction\\}
Majorana zero modes (MZMs) are charge-neutral non-Abelian quasiparticles~\cite{Kitaev1,Kitaev2}. 
They have attracted much interest because of their potential application in fault-tolerant topological quantum computing~\cite{Kitaev2,Sarma,Nayak,Shnirman}.
The occurrence of zero-bias peaks in tunneling spectroscopy is one of the  experimental signatures of MZMs~\cite{Law}. 
A promising platform to realize MZMs are semiconductor nanowires proximitized to superconductors~\cite{Sau}, and  
{\ {magnet-superconductor hybrid systems~\cite{Crawford,Wong,Huang,Mascot}}}, where MZMs  are expected to develop at 
both ends of the wire. {\ {However, in 
these materials the so-called ``sweet spot'' in couplings, where Kitaev theory is primarily developed, is difficult to reach}. Recently, the realization of MZMs was also proposed in quantum-dot-superconductor linear arrays~\cite{Jay,Souto,Mills}.
These quantum-dots systems~\cite{Leijnse,Bara,Csonka,Deng,Liu,Loss} are expected to reduce the problem of random-disorder potential, 
as compared to the proximitized semiconductor nanowires where the 
effect of disorder is strong~\cite{Jay}
and may create false signals of MZMs in tunneling spectra~\cite{Stanescu}. 
Interestingly, the experimental  realization of a minimal Kitaev chain  has been demonstrated using two quantum dots coupled through a short
superconducting-semiconductor hybrid (InSb nanowire)~\cite{Dvir}. Remarkably, in this experiment~\cite{Dvir}, 
two localized MZMs were observed in tunneling
conductance measurements at the sweet spot $t_h=\Delta$. 
{\ {Very recently the experimental realization of  a three-site Kitaev chain geometry in
 a quantum-dot system has been reported~\cite{Bordin}, opening the doors towards the realization of  
longer Kitaev chains using many coupled quantum dots.}} 

In topological quantum computation, it is required to move and perform braiding operations of the MZMs~\cite{Alicea,Aasen}.  
A strict 1D geometry is not sufficient to perform  such braiding operations, 
because in 1D the MZMs can fuse during their exchange process~\cite{Pandey,Han}. 
 To realize non-Abelian statistics (or braiding), $T$ and $Y$-shaped wires geometry have been 
proposed~\cite{Alicea,Heck,Sekania, Harper,Giuliano}. 
It has been shown that the MZMs in $T$-shape nanowires can be transformed 
under exchange, similarly to 2D $p+ip$ superconducting systems, displaying non-Abelian statistics~\cite{Alicea}.
Braiding-based gates with MZMs using quantum dot arrays were recently proposed~\cite{Boross}.

{\ {This paper focuses on finding MZM modes near the junction of 
interacting $Y$-shaped  quantum dots near the sweet-spot ($t_h$=$\Delta$), important for braiding and
for demonstrating the non-Abelian statistics of MZMs.
However, in the context of proximity-induced semiconductor nanowires 
(in the limit of $\Delta$ $\ll$ $t_h$), there are only a few studies 
related to ground states of $T$-shaped or  $Y$-shaped wires~\cite{Zhou,Ardone,deb}.}} The sub-gap properties of a three-terminal Josephson junction (composed of effective spinless $p$-wave superconductors),
 joined into a $T$-shaped normal-metallic region, has been studied using the scattering matrix approach in the non-interacting limit~\cite{Ardone}. 
They found that depending upon the superconducting phase of each arm, 
the  Majorana zero mode extended into the  metallic region  of either all three legs or two legs of the $T$-shape wire.
{\ {In Ref.~\cite{Khanna} the existence of a zero-energy parafermionic mode was
proposed in multi-legged star junctions of quantum Hall states.}}
However, in these studies the precise form of the Majorana wave functions and the effect of Coulomb interactions were not addressed~\cite{Zhou,Ardone}. 
The repulsive Coulomb interaction is expected to suppress the pairing-induced bulk-gap and can affect the stability of Majorana modes~\cite{Oleg,Martin}.

Motivated by the above described 
recent progress in the realization of minimal Kitaev chains using quantum dots~\cite{Dvir,Bordin},
we study the Majorana zero modes in $Y$-shaped interacting Kitaev wires.
We address the exact form of Majorana wavefunctions near the junction and its dependency on 
the superconducting phases at each wire. Assuming each arm of the $Y$-shape wire
can take different values for the superconducting (SC) phase (see Fig.~\ref{fig1}{\bf a}), 
we solve {\it exactly} the Kitaev model for $Y$-shapes wires working at the sweet spot $t_h=\Delta$. 
Remarkably, in terms of Majorana operators, we are able to write four independent commuting Hamiltonians, 
consisting of the three arms ($I, II, III$) and one central region ($IV$), as shown in Fig.~\ref{fig1}{\bf b}.
This allow us to diagonalize the full system independently for each region 
and solve {\it exactly} at the sweet spot. 
We find the expected three single-site localized  MZMs on the edge sites of the  $Y$-shape wire.
Surprisingly, depending on the SC phase values, {\ {we also 
find exotic multi-site MZMs near the central region,
not bounded to be on only one site as in the Kitaev model~\cite{Nagae}. 
Instead they are located on two or three edge sites of the different legs of
the $Y$-shape wire. Note that this is conceptually totally different from the standard exponential decay
of the Majorana wave function away from the sweet spot or under non-ideal conditions. The multi-site nature
of the MZMs unveiled here is an intrinsic, previously unknown, property of the $Y$-geometry system
at the sweet spot.}}  


\begin{figure}[!ht]
\hspace*{-0.5cm}
\vspace*{0cm}
\begin{overpic}[width=1.0\columnwidth]{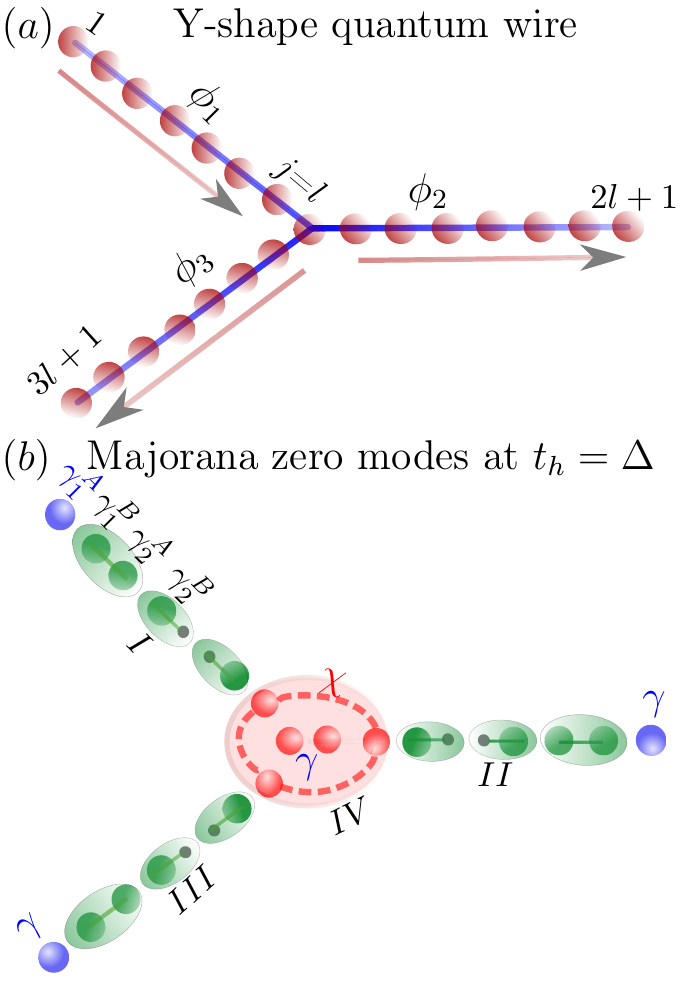}
\end{overpic}
\caption{(a) Schematic representation of the $Y$-shape Kitaev wires used here. $\phi_1$, $\phi_2$, and $\phi_3$ are the phases of the $p$-wave superconductor
in each arm of the $Y$-shape wire. Arrows denote the directions 
of the pairing terms $\Delta$ and site index $j$  in each wire.
(b) Pictorial representation of the Majorana zero modes in the $Y$-shape Kitaev wire unveiled here, at $t_h=\Delta$. 
There are always three localized MZMs $\gamma$ (purple color) at each end of the $Y$-geometry. 
Depending upon $\phi_1$, $\phi_2$ and $\phi_3$ for each arm, near the wire's junction 
we observed only two situations: either (i) one multi-site MZM ($\chi$) or (ii) two multi-site MZMs ($\chi$) and one single-site MZM ($\gamma$).
Green color  MZMs form ordinary fermions at each bond.}
\label{fig1}
\end{figure}

Furthermore, we perform much needed unbiased simulations of the Majorana zero modes in $Y$-shaped 
interacting Kitaev wires,
using density matrix renormalization group (DMRG)~\cite{alvarez,Nocera}.
In the non-interacting limit, we find peaks in the site-dependent local density-of-states (LDOS)
at the exact locations predicted by our analytical calculations.  
In order to compare the stability of single-site vs. multi-site MZMs,
 we examine the electron and hole components of the LDOS separately~\cite{Herbrych,Rachel}, 
against the increase in repulsive Coulomb interactions.
Interestingly, the LDOS$(\omega,j)$ calculations indicate the multi-site MZMs          
are as equally stable as the single-site MZMs residing at the ends of the $Y$-shaped wire.

\noindent {\small \bf Model Hamiltonian\\}
{\ {The Hamiltonian for the $Y$-shaped Kitaev model, with superconducting phases $\phi_1$, $\phi_2$, and $\phi_3$ at each arm, can be divided into four different parts.
The Hamiltonian for each leg can be written as: 
\begin{eqnarray}
        H^I = \sum_{j=1}^{l-1}\left( -t_hc^{\ {\dagger}}_{j}c_{j+1} + e^{i\phi_1} \Delta  c_j c_{j+1} + H.c. \right),\\
        H^{II} = \sum_{j=l+2}^{2l}\left( -t_hc^{\ {\dagger}}_{j}c_{j+1} + e^{i\phi_2} \Delta  c_j c_{j+1} + H.c. \right),\\
        H^{III} = \sum_{j=2l+2}^{3l}\left( -t_hc^{\ {\dagger}}_{j}c_{j+1} + e^{i\phi_3} \Delta  c_j c_{j+1} + H.c. \right).
\end{eqnarray}
Moreover, the Hamiltonian for the central site $l+1$ joining each leg edge site can be written as: 
\begin{eqnarray}
        H^{IV} = \left( -t_hc^{\ {\dagger}}_{l}c^{\phantom \dagger}_{l+1} +e^{i\phi_1} \Delta  c_l c_{l+1} + H.c. \right) \nonumber\\
                 \left( -t_hc^{\ {\dagger}}_{l+1}c^{\phantom \dagger}_{l+2} +  e^{i\phi_2} \Delta  c_{l+1} c_{l+2} + H.c. \right) \nonumber\\
             \left( -t_hc^{\ {\dagger}}_{l+1}c^{\phantom \dagger}_{2l+2} + e^{i\phi_3} \Delta  c_{l+1} c_{2l+2} + H.c. \right).
\end{eqnarray}
}}
\noindent {\bf \\Results\\}
\noindent {\small \bf \\Description of analytical method\\}
{\ {We noticed that the $Y$-shape Kitaev model can be solved analytically 
at the sweet spot $t_h=\Delta$ and $V=0$}, and we will focus on 
three different sets of superconducting phases: (i) $\phi_1=\pi$, $\phi_2=0$, and  $\phi_3=0$,
(ii) $\phi_1=0$, $\phi_2=0$, and  $\phi_3=0$, and (iii) $\phi_1=0$, $\phi_2=0$, and  $\phi_3=\pi/2$.
First, we divide the system Hamiltonian into four different parts 
(for details see Model Hamiltonian) in terms of spinless fermionic operators. 
Then, we rewrite the system Hamiltonian in terms of Majorana operators, 
using the transformation $c_j=\frac{1}{\sqrt{2}} e^{-i\frac{\phi_k}{2}}\left(\gamma^A_{j}+i\gamma^B_{j}\right)$~\cite{Kitaev1}, where $k$ refers to each of the legs.
In terms of the Majorana operators, remarkably 
the system can be written as four independent {\it commuting} Hamiltonians: 
(1) the three independent 1D wires ($I, II, III$) (see Fig.~\ref{fig1}{\bf b})  
and (2) the central region ($IV$), consisting only of five Majorana operators 
(two from the central site and three from the near edge sites of each leg), 
as shown in Fig.~\ref{fig1}{\bf b}.
As already expressed, this procedure  allows us to solve the $Y$-shape 
Kitaev model {\it exactly} at the sweet spot $t_h=\Delta$ and $V=0$,
for any values of the SC phases of each arm. 
For all three cases of SC phases discussed above, we find there is one Majorana zero mode 
at the outer edge (single-site) of each arm (purple color in  Fig.~\ref{fig1}{\bf b}),
 as expected intuitively.
{\ {The main novelty is that depending on the phase values of each arm, in addition to the above mentioned outer edge MZMs, we find at the central region either
 (i) only one multi-site MZM ($\chi$) or (ii) two multi-site MZMs ($\chi$) accompanied with 
one single-site MZM ($\gamma$).
These multi-site MZMs ($\chi$) are located at the edge sites (near the central region)  of each arm in the crucial limit of $t_h$=$\Delta$ that allows for the analytic solution.}}
These exotic multi-site MZMs could be realized in quantum dot experiments, 
by changing the phases of each arm in a $Y$-shape geometry of quantum dots arrays.
  
In terms of Majorana operators, the Hamiltonian of each leg is independent of the SC phases ($\phi_1, \phi_2,\phi_3$). 
Using the transformations $c^I_j= \frac{1}{\sqrt{2}} e^{-i\phi_1/2}\left(\gamma^I_{A,j}+i\gamma^{I}_{B,j}\right)$, $c^{II}_j= \frac{1}{\sqrt{2}}e^{-i\phi_2/2}\left(\gamma^{II}_{A,j}+i\gamma^{II}_{B,j}\right)$ and $c^{III}_j= \frac{1}{\sqrt{2}}e^{-i\phi_3/2}\left(\gamma^{III}_{A,j}+i\gamma^{III}_{B,j}\right)$, the Hamiltonian for the three legs $H^{I}, H^{II}$ and $H^{III}$ can be written in terms of Majorana
operators as:
\begin{eqnarray}
 H^I =-2i\Delta \sum_{j=1}^{l-1}\left( \gamma^{I}_{A,j+1}\gamma^{I}_{B,j}\right),\\
  H^{II} =-2i\Delta \sum_{j=l+2}^{2l}\left( \gamma^{II}_{A,j+1}\gamma^{II}_{B,j}\right),\\
 H^{III} =-2i\Delta \sum_{j=2l+2}^{3l}\left(\gamma^{III}_{A,j+1}\gamma^{III}_{B,j}\right),
\end{eqnarray}
where we have used the sweet-spot property $t_h=\Delta$. 
In these equations, the Majorana operators $\gamma^{I}_{A,1}$, $\gamma^{II}_{B,2l+1}$, and $\gamma^{III}_{B,3l+1}$
are absent~\cite{Alicea}, and commute with these Hamiltonians, 
which indicates the presence of three single-site MZMs at the edge sites of the $Y$-shape Kitaev wire 
(see Fig.~\ref{fig1}{\bf b}), similarly as in the original Kitaev chain exact solution.
For the central region, the Hamiltonian $H^{IV}$ depends upon the SC phases. To illustrate the physics
unveiled here, we solve  $H^{IV}$ for  
three different cases, in order to understand the nature of the central MZMs.

\begin{figure*}[!ht]
\hspace*{-0.5cm}
\vspace*{0cm}
\begin{overpic}[width=2.0\columnwidth]{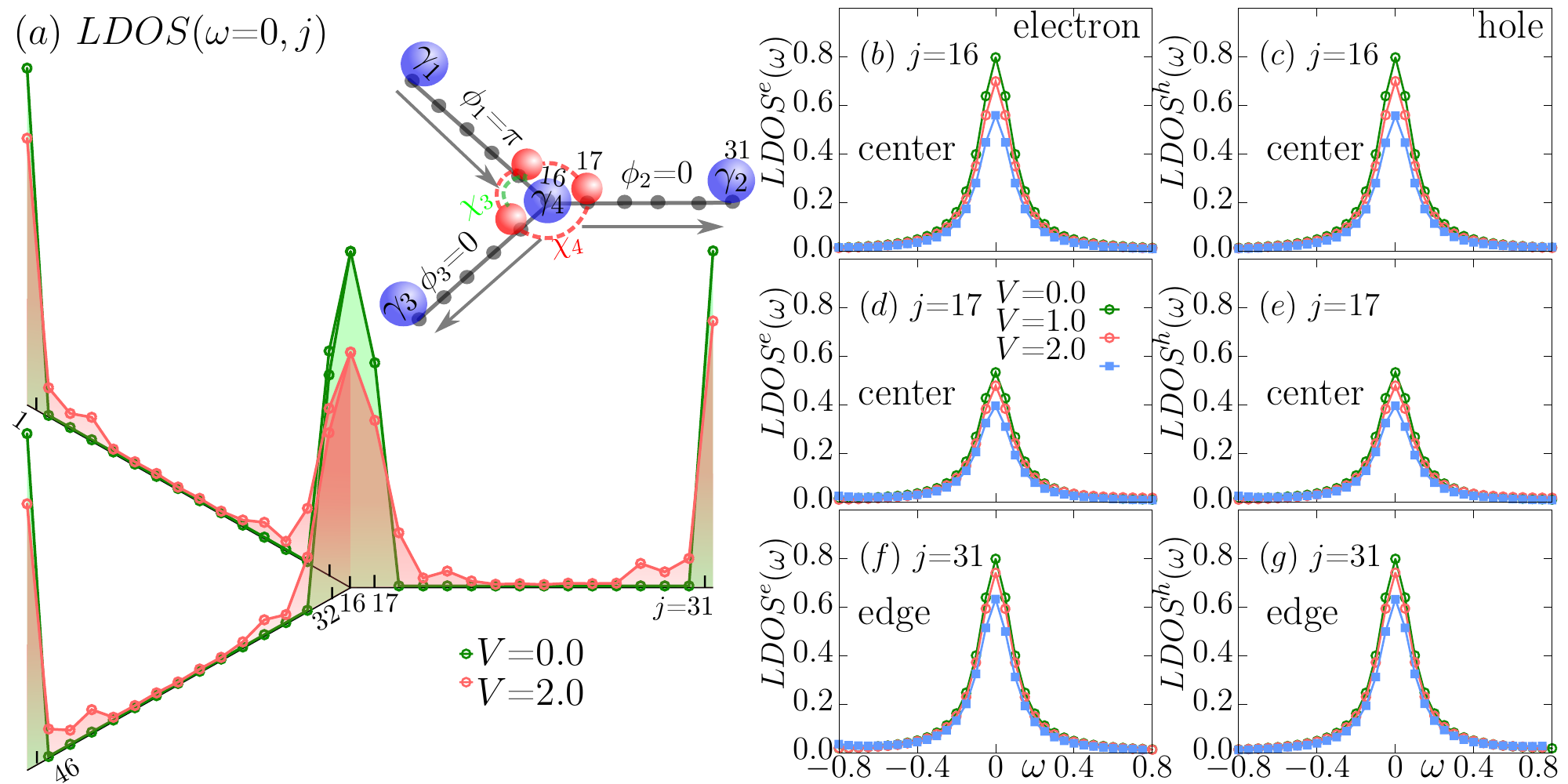}
\end{overpic}
\caption{Schematic representation of  Majorana zero modes in a $Y$-shape Kitaev wire, for the case $\phi_1=\pi$,  $\phi_2=0$, and  $\phi_3=0$, at $t_h=\Delta=1$.
(a) There are a total of six MZMs. For $V=0$, three single-site MZMs ($\gamma_1$, $\gamma_2$, and $\gamma_3$) 
at the edge sites $j=1$, $31$, $46$ and in addition one single-site MZM ($\gamma_4$) at the central site $j=16$. Near this central region, there are two  additional multi-site Majoranas:  $\chi_3$ (distributed on sites $j=15$ and $32$) and $\chi_4$ (distributed on sites $j=15$, $17$, and $32$), which leads to a spectral weight $2/3$ (compared to the single-site localized MZMs).
At robust repulsion $V=2$ the Majoranas remain exponentially localized over a few sites.
The electron and hole parts of the LDOS$(\omega)$ vs. $\omega$ are shown for sites 
(b) $j=1$, (d) $j=17$, and (f) $j=31$.
For $V \leq 2$, the electron and hole parts of the LDOS$(\omega)$ are almost equal and shows peaks at $\omega=0$, indicating the presence of stable MZMs on these sites.
}
\label{fig2}
\end{figure*}

\noindent {\bf \\ The case $ {\bf{\phi_1=\pi}}$, ${\bf{\phi_2=0}}$, and  ${\bf{\phi_3=0}}$\\}
 To explain the exact solution of Majorana wave functions in $Y$-shape geometries, 
we start with phase values $\phi_1=\pi$, $\phi_2=0$, and  $\phi_3=0$ on arms $I,$ $II,$ and $III$, respectively. 
 For these phase values, the pairing term at each arm preserves the rotational symmetry of the system,
 as under 120$^{\circ}$ rotation  around the central sites $l+1$, the system Hamiltonian remains invariant due to the $\pi$ phase in arm $I$ 
(see SM for more detail~\cite{SM})~\cite{Dagotto1,Dagotto2}.

For the central sites, using the relations  $c_l=\frac{1}{\sqrt{2}}  e^{-i\phi_1/2}\left(\gamma^{I}_{A,l}+i \gamma^{I}_{B,l}\right)$,
$c_{l+1}=\frac{1}{\sqrt{2}} \left(\gamma^{IV}_{A,l+1}+i \gamma^{IV}_{B,l+1}\right)$, $c_{l+2}=\frac{1}{\sqrt{2}} e^{-i\phi_2/2} \left(\gamma^{II}_{A,l+2}+i \gamma^{II}_{B,l+2}\right)$, and
$c_{2l+2}=\frac{1}{\sqrt{2}}  e^{-i\phi_3/2} \left(\gamma^{III}_{A,2l+2}+i \gamma^{III}_{B,2l+2}\right)$, 
the sector $H^{IV}$ (with $\phi_1=\pi$,  $\phi_2=0$, and  $\phi_3=0$) can be transformed in terms of Majorana operators as:
\begin{eqnarray}
 H^{IV}=-2i\Delta \left( \gamma^{I}_{B,l} + \gamma^{II}_{A,l+2} + \gamma^{III}_{A,2l+2}\right) \gamma^{IV}_{B,l+1}.
\end{eqnarray}
Note that in Eq.~8 the Majorana operator $\gamma^{IV}_{A,l+1}$  at the central
 site $l+1$ is absent, signaling the presence of a single-site MZM at site  $j=16$ (for a $L=46$ sites system).  Next, we write Eq.~8 in terms
of  a $4\times4$ matrix in the  basis of $\left(\gamma^{I}_{B,l},\gamma^{II}_{A,l+2},  \gamma^{III}_{A,2l+2}, \gamma^{IV}_{B,l+1}   \right)$
and obtained four eigenvalues ($-\sqrt{3}, \sqrt{3}, 0, 0$). The last two eigenvalues ($e_3$, $e_4$) 
are zero denoting MZMs, and the associated eigenvectors $\chi_3= -\frac{1}{\sqrt{2}}  \gamma^{I}_{B,l}+ \frac{1}{\sqrt{2}} \gamma^{III}_{A,2l+2} $
 and $\chi_4=-\frac{1}{\sqrt{6}}\gamma^{I}_{B,l} + \sqrt{\frac{2}{3}}\gamma^{II}_{A,l+2} - \frac{1}{\sqrt{6}} \gamma^{III}_{A,2l+2}$ emerge, with properties $\chi^{\phantom \dagger}_3=\chi^{\dagger}_3$ and $\chi^{\phantom \dagger}_4=\chi^{{\dagger}}_4$. {\ {In addition,
$\chi_3$ and $\chi_4$ also commute with $H^{IV}$, and its amplitude splits 
into multiple sites (i.e. not bounded at only one site); 
these properties confirm the presence of two multi-site MZMs in the central region. 
The MZM  $\chi_3$ is located on two sites $j=15$ and $j=32$ with equal amplitude, whereas
 the MZM  $\chi_4$ is distributed  on three sites $j=15, 17$ and $32$
with different amplitudes (see Fig.~\ref{fig2}{\bf a}). Note that this is unrelated to exponential
decaying wave functions, as it occurs away from the sweet spot. There is no ``tail'' in these multi-site wave
functions but instead the core is fully spread into a relatively small number of sites.}} 
The diagonalized Hamiltonian for the central site can be written as
\begin{equation}
H^{IV}= 2\sqrt{3}\Delta\left({\chi_2}^{{\dagger}} \chi_2-\frac{1}{2} \right), 
\end{equation}
where  $\chi_2=\frac{i}{\sqrt{6}} \gamma^{I}_{B,l} + \frac{i}{\sqrt{6}} \gamma^{II}_{A,l+2} + \frac{i}{\sqrt{6}} \gamma^{III}_{A,2l+2}+ \frac{1}{\sqrt{2}} \gamma^{IV}_{B,l+1}$ is an ordinary fermion.

The other remaining  Hamiltonians for each arm can be diagonalized using fermionic operators
$d_{k,j}=\frac{1}{\sqrt{2}} \left(\gamma^{k}_{B,j}+i \gamma^{k}_{A,j+1} \right)$,
where $k=I, II$, or $III$. 
The diagonalized  Hamiltonian $H^{I}$,  $H^{II}$, and  $H^{III}$ are:
\begin{equation}
H^{I}=2\Delta \sum_{j=1}^{l-1}\left( d^{\ {\dagger}}_{I,j} d_{I,j}-\frac{1}{2}\right),\\ 
\end{equation}
\begin{equation}
 H^{II}= 2\Delta \sum_{j=l+2}^{2l} \left( d^{\ {\dagger}}_{II,j} d_{II,j}-\frac{1}{2}\right),
\end{equation}
\begin{equation}
 H^{III}= 2\Delta \sum_{j=2l+2}^{3l} \left( d^{\ {\dagger}}_{III,j} d_{III,j}-\frac{1}{2}\right).
\end{equation}
In conclusion, our analytical calculations predict a total of six MZMs, a surprisingly large number, 
for the case of $\phi_1=\pi$, $\phi_2=0$ and $\phi_3=0$.
These six MZMs lead to eight fold-degeneracy in the ground state of the system, after including the
result of Majorana fusion,~\cite{Nayak} 
which we also find is fully consistent with our numerical Lanczos calculations
(see SM for more details~\cite{SM}).

\begin{figure*}[!ht]
\hspace*{-0.5cm}
\vspace*{0cm}
\begin{overpic}[width=2.0\columnwidth]{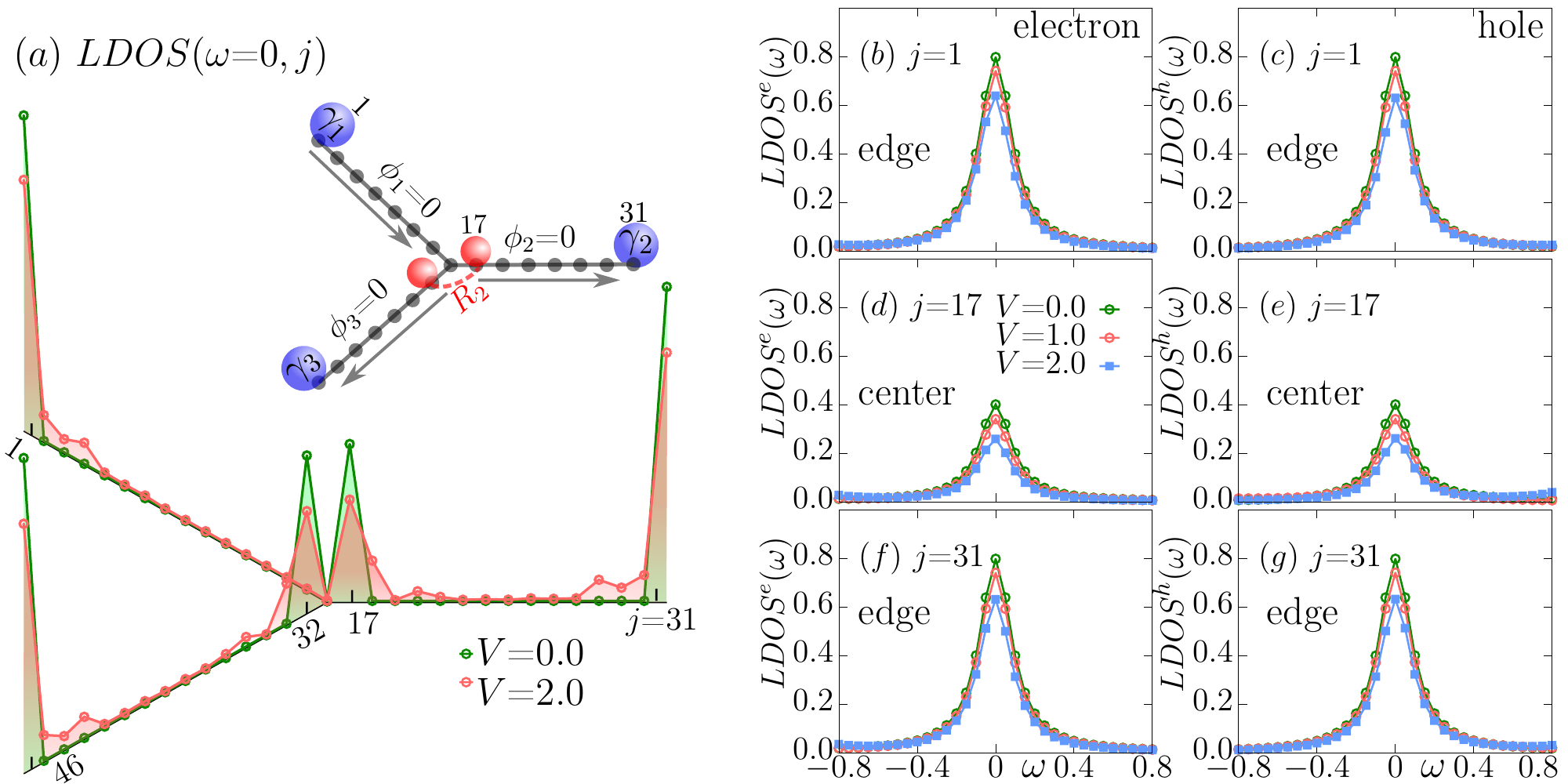}
\end{overpic}
\caption{Schematic representation of  Majorana zero modes in a $Y$-shape Kitaev wire, with $\phi_1=0$,  $\phi_2=0$, and  $\phi_3=0$,
and at $t_h=\Delta=1$.  
(a) In this case, there are a total of four MZMs. 
Three are localized at the end sites $j=1$, $31$, and $46$. Near the central region,
 there is one extra multi-site MZM $\chi_4$. 
The multi-site MZM $R_2$  is equally distributed among the central sites $j=17$ and $32$, leading to an 
spectral weight exactly $1/2$ (compared to the localized MZMs with weight 1)
as shown in the LDOS$(\omega=0,j)$ on these central sites. For $V=2$, the MZMs are spread over a few sites compared to the $V=0$ case.
The electron and hole parts of the LDOS$(\omega)$ vs. $\omega$ are shown for sites (b) $j=1$, (d) $j=17$, and (f) $j=31$.
For $V \leq 2$ the electron and hole parts of the LDOS$(\omega)$ are almost equal and shows peaks at $\omega=0$,
indicating the presence of stable MZMs on  sites $j=1,17,31,32$, and $46$. Note the weight at $j=17$ is half that
of sites $j=1$ and $j=31$.
}
\label{fig3}
\end{figure*}

Next, we analyze the  stability of MZMs in the presence of a 
nearest-neighbor interaction $H_I=Vn_jn_{j+1}$, where $n_j$ is the electronic density at site $j$.
We calculate the LDOS,  using DMRG for a $Y$-shaped geometry with system size $L=46$.
As shown in  Fig.~\ref{fig2}{\bf a}, the site dependent LDOS($\omega=0,j$) shows sharp peaks for the  
edge sites $j=1$, $31$, and $46$, indicating three single-site localized MZMs at each edge of the arms.
At the central site $l+1$, there is a sharp peak with same height as for the edge sites,
 showing the presence of a single-site MZM $\gamma_4$ at site $j=16$, as already discussed. 
Interestingly, there are three other
peaks in that LDOS($\omega=0,j$) on sites $j=15, 17,$ and $32$. 
From the expressions of $\chi_3$ and $\chi_4$
presented in the previous page, it can be deduced that the height should be exactly 2/3 for sites $j=15, 17,$ and $32$ as compared to the edge sites.
These three peaks signal the presence of two multi-site MZMs, distributed 
over three central sites ($j=15, 17$, and $32$). 

Increasing the interaction strength to $V=2$, the Majoranas are no longer strictly localized at a single site $j$.
Now the MZMs are decaying over a few more sites and consequently the peak height of the
LDOS($\omega=0,j$) decreases  (Fig.~\ref{fig2}{\bf a}).
This shows that the Majorana zero modes are topologically protected against moderate values of the Coulomb interaction.

To compare the topological protection against $V$, for single-site  and multi-site MZMs, 
 we calculate the electron and hole parts of LDOS($\omega,j$), separately for the 
edge and central sites (right panel of  Fig.~\ref{fig2}).
In  Figs.~\ref{fig2}{\bf b} and {\bf c}, we show the electron and hole 
part of LDOS($\omega,j$) for the central site $j=16$. With increasing $V$, the peak height
of the electron and hole portions of the LDOS($\omega$) at $\omega=0$ decrease to  
the same values, showing the preservation of its MZM nature ($\gamma=\gamma^{\ {\dagger}}$)~\cite{Herbrych}.  
Due to the rotational symmetry of the system, 
sites near the center $j=15,17$, and $32$ are equivalent and they
behave very similarly increasing $V$. In  Figs.~\ref{fig1}{\bf d} and {\bf e}, 
we show  the electron and hole parts of  LDOS($\omega,j$) 
for site $j=17$.  As discussed previously, the two multi-site MZMs $\chi_3$ and $\chi_4$ are distributed on sites ($j=15$ and $32$) and ($j=15, 17$, and $32$),
with total amplitude 2/3 on each site ($j=15, 17$, and $32$), 
which leads to  a spectral weight $2/3$ (compared to the single-site MZMs with weight 1) 
in the LDOS($\omega,j$) for site $j=17$ (also for $j=15$ and $32$ at $V=0$).
  Figures~\ref{fig2}{\bf f} and {\bf g} show the electron and hole part of  LDOS($\omega,j$) for the edge site $j=31$ with increasing $V$ (note: sites $j=1$ and $46$ are equivalent). The rate of decrease in peak height in electron and hole part of LDOS($\omega,j$), 
for multi-site MZMs at site $j=17$ and single-site MZM at edge site $j=31$ are almost identical when increasing $V$ (in the range $V\le 2$).
The LDOS($\omega,j$) of the local MZM at the central site $j=16$ decreases with 
a slightly faster rate because it develops a finite overlap with $\chi_3$ and $\chi_4$, with increasing $V$.

\noindent {\small \bf \\ The case $ {\bf{\phi_1=0}}$, ${\bf{\phi_2=0}}$, and  ${\bf{\phi_3=0}}$\\}
Let us consider now the same $Y$-shape geometry but with the same phase $\phi=0$ on each arm.
Surprisingly, for  the $\phi_1=0$, $\phi_2=0$, and  $\phi_3=0$ case, the pairing term 
in the Hamiltonian breaks rotational symmetry,
as after a 120$^{\circ}$ anti-clockwise rotation  around the central sites $l+1$, 
the pairing term in leg $I$ changes its sign
 (becomes negative due fermionic anticommutations) (see also SM for details~\cite{SM}).
Arms $II$ and $III$ have reflection symmetry around the central site $l+1$.
Using the same transformations as previously discussed, the $H^I$, $H^{II}$ and $H^{III}$ 
of each arm in terms of Majorana 
operators can be written in similar form as described using Eqs.~5, 6, and 7.
 Again, in these equations, the Majorana operators $\gamma^{I}_{A,1}$, $\gamma^{II}_{B,2l+1}$, and $\gamma^{III}_{B,3l+1}$
are absent, indicating the presence of three edge single-site MZMs at sites $j=1,31$, 
and $46$ of a $Y$-junction with a total of 46 sites. The Hamiltonian for the central 
region $H^{IV}$ can be transformed as:
\begin{eqnarray}
 H^{IV} = -2i\Delta \left[\gamma^{IV}_{A,l+1} \gamma^{I}_{B,l} +\left( \gamma^{II}_{A,l+2} + \gamma^{III}_{A,2l+2}\right) \gamma^{IV}_{B,l+1} \right]. \nonumber
\end{eqnarray}
In the above equation, $H^{IV}$ has a reflection symmetry  [$\gamma^{II}_{A,l+2} \leftrightarrow \gamma^{III}_{A,2l+2}$].
Thus, defining the operators
\begin{eqnarray}
        R_1 = \frac{1}{\sqrt{2}}\left(\gamma^{II}_{A,l+2} +\gamma^{III}_{A,2l+2}\right),\nonumber   \\ 
        R_2 = \frac{1}{\sqrt{2}}\left(\gamma^{II}_{A,l+2} -\gamma^{III}_{A,2l+2}\right),\ \     
\end{eqnarray}
the Hamiltonian  $H^{IV}$ further simplifies as
\begin{equation}
 H^{IV} = -2i\Delta \gamma^{IV}_{A,l+1} \gamma^{I}_{B,l}  - 2\sqrt{2}i \Delta R_1 \gamma^{IV}_{B,l+1}.
\end{equation}

In Eq.~14 the operator $R_2$ is absent and has the properties $R^{\phantom \dagger}_2=R^{\ {\dagger}}_2$, and $[H^{IV},R_2]=0$, 
indicating $R_2$ is a Majorana zero mode.
The  Majorana zero mode $R_2 = \frac{1}{\sqrt{2}}\left(\gamma^{II}_{A,l+2} -\gamma^{III}_{A,2l+2}\right)$ is equally distributed on
sites $17$ ($II$ leg) and $32$ ($III$ leg) (see Fig.~\ref{fig3}{\bf a}), showing that $R_2$ {\ {is indeed a multi-site MZM}}.
Next, we write Eq.~14  in terms of a  $4\times4$ matrix in the basis of $\left[\gamma^{IV}_{A,l+1},R_1,\gamma^{IV}_{B,l+1},  \gamma^{I}_{B,l} \right]$
and obtain four eigenvalues ($-\sqrt{2}, \sqrt{2}, -1, 1$). 
The central region diagonal Hamiltonian can be written in terms of ordinary fermions $\chi_2$ and $\chi_4$ as
\begin{equation}
H^{IV}= 2\sqrt{2}\Delta\left({\chi_2}^{{\dagger}} \chi_2-\frac{1}{2} \right) +2\Delta\left({\chi_4}^{{\dagger}} \chi_4-\frac{1}{2} \right), 
\end{equation}
with $\chi_2=\frac{1}{\sqrt{2}} \left(i R_1+ \gamma^{IV}_{B,l+1}\right)$ and $\chi_4=\frac{1}{\sqrt{2}} \left(i \gamma^{IV}_{B,l+1} + \gamma^{I}_{B,l}\right)$
(see SM for more detail~\cite{SM}).
The diagonalized  Hamiltonian $H^{I}$,  $H^{II}$, and  $H^{III}$ takes similar form as Eqs.~10, 11, and 12, 
in terms of the ordinary $d_{k,j}$ fermionic 
operators. In summary, our analytical calculation finds a total of four MZMs (three single-site at the edge sites and one multi-site MZM near the center region).
These four MZMs results in four-fold degeneracy in the ground state of the system, which is also consistent with our full-diagonalization numerical results.

 Figure~\ref{fig3} shows DMRG results for $L=46$ sites at $t_h=\Delta$ and for different values of the Coulomb interaction $V$. 
At $V=0$, the site dependent LDOS($\omega=0,j$) shows sharp localized peaks at the edge sites $j=1, 31,$ and $46$,
indicating three single-site MZMs on those edge sites, as expected. 
Near the center,  the LDOS($\omega=0,j$) displays two peaks  at sites $j=17$ and $32$,
with height $1/2$ compared to the edge sites, suggesting the presence of a multi-site MZM.
 With increase in interaction ($V=2$), these MZMs remain exponentially localized over a few sites (Fig.~\ref{fig3}{\bf a}). 
To compare the stability of single-site and multi-site MZMs,
we calculate the electron and hole parts of LDOS($\omega,j$) for different values $V$.
 The peak values for LDOS$^e(\omega)$ (Fig.~\ref{fig3}{\bf b} and {\bf f}) and  LDOS$^h(\omega)$
(Fig.~\ref{fig3}{\bf c} and {\bf g}) at $\omega=0$, for edge sites $j=1$ and $31$, decrease to the same values with increase in $V$. This
shows that the characteristic features of the single-site MZM remain and
the spectral weight of electron and hole part of  LDOS($\omega,j$) 
are equal~\cite{Herbrych}, at moderate values of $V\le2$. 

Interestingly, the spectral weight of electron and hole of LDOS($\omega$) for site $j=17$,
takes value half (compared to the single-site MZMs on edge sites) at $V=0$.
 This is because the  multi-site MZM $R_2= \frac{1}{\sqrt{2}}\left(\gamma^{II}_{A,17} -\gamma^{III}_{A,32}\right)$ is 
equally distributed at sites $j=17$ and $32$.
Increasing the repulsion strength $V$, the peak values of  LDOS$^e(\omega)$ (Fig.~\ref{fig3}{\bf d}) and  LDOS$^h(\omega)$ (Fig.~\ref{fig3}{\bf e}) 
are reduced (but still take the same values). The rate of decrease in peak height for single- and multi-site MZMs are almost the same,
which suggests these Majorana modes are equally topologically protected against $V$.

{\ {Figure~\ref{fig4} shows the numerical Bogoliuvov-de Gennes (BdG)~\cite{Kitaev1,Kitaev2} results for $L=46$ sites at $V=0$ and for different values of the pairing amplitude $\Delta$, to analyze what occurs away from the sweet spot.
At $\Delta=t_h=1$, the three edge MZMs are single-site localized on one end site of each leg, and do not decay exponentially, whereas the only multi-site MZM ($R_2$),
is localized at sites $j=17$ and $32$ (with equal peak height in  LDOS($\omega,j$)). At $\Delta/t_h=0.6$ the four  MZMs are quite stable and exponentially 
decay to a few sites.  On the other hand, for even smaller values 
of $\Delta/t_h$, such as 0.2, the edge sites MZMs decay exponentially over many sites. 
Note that the multi-site Majorana also  decays exponentially over 
many sites of arms  $II$ and $III$. For that reason, 
for $\Delta=0.2$, the multi-site Majorana ($R_2$) overlaps with the edge MZMs ($\gamma_2$ and $\gamma_2$) on arms $II$ and $III$ (see Fig~\ref{fig4}).      
The overlap of central MZM to edge MZMs could give rise to interesting features in electron transport experiments~\cite{deb}.}}
\begin{figure}
\centering
\includegraphics[width=0.48\textwidth]{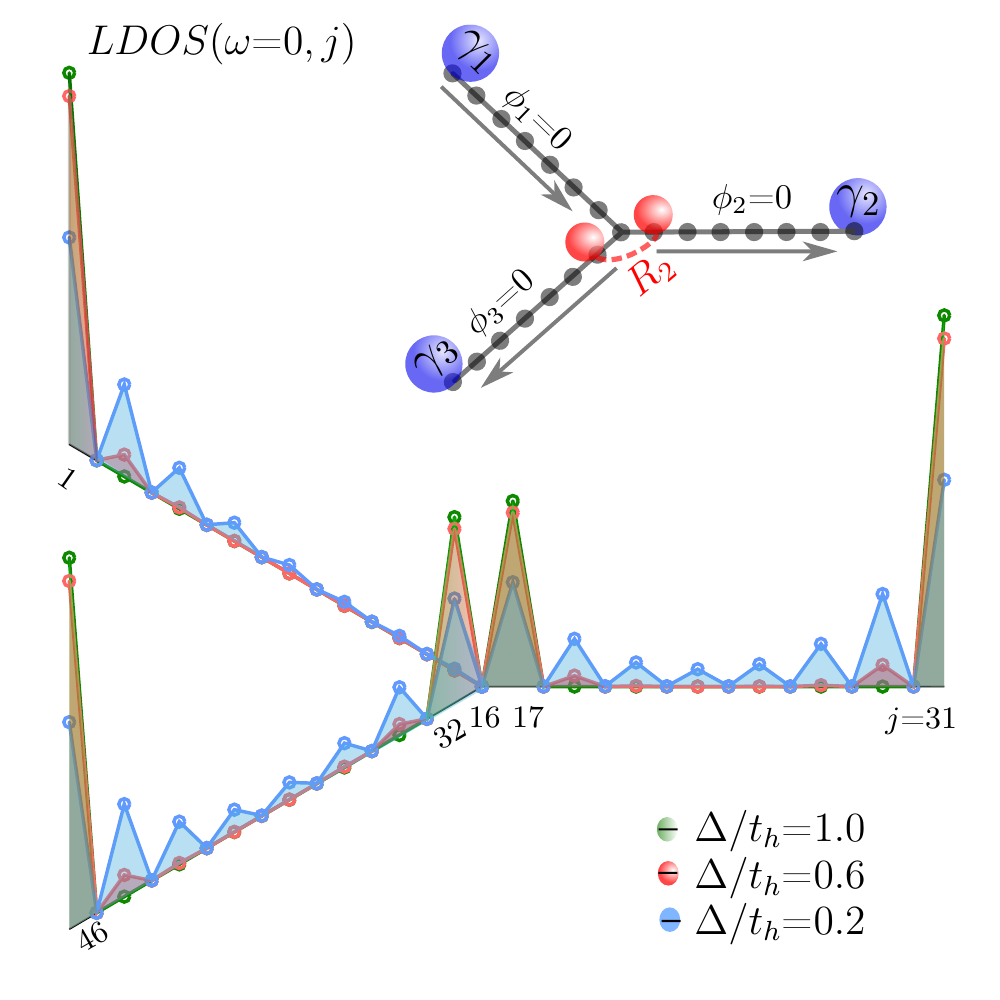}
\caption{{\ {Schematic representation of  Majorana zero modes in the $Y$-shape Kitaev wire with $\phi_1=0$,  $\phi_2=0$, and  $\phi_3=0$. 
The LDOS($\omega=0,j$) vs site $j$ for different values of $\Delta/t_h$.
For $\Delta/t_h=1.0$ and $0.6$, the three single-site  MZMs are almost localized 
at the end sites $j=1$, $31$, and $46$, while near the central region, 
the multi-site Majorana $\chi_5$ is located on sites $j=17$ and $32$.
For  $\Delta/t_h=0.2$, the central MZM ($R_2$) overlaps with the edge MZMs ($\gamma_2$ and $\gamma_3$). 
}}}
\label{fig4}
\end{figure}

\begin{figure}
\centering
\includegraphics[width=0.48\textwidth]{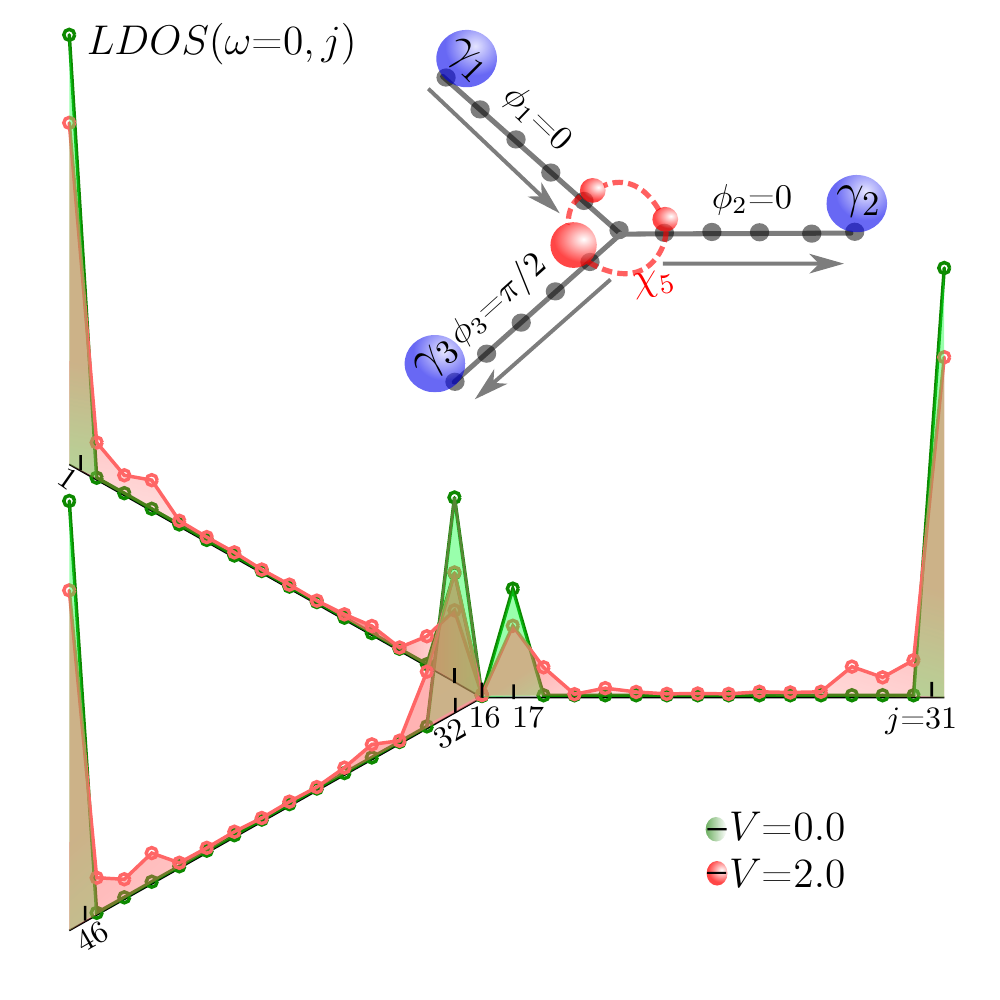}
\caption{Schematic representation of  Majorana zero modes in the $Y$-shape Kitaev wire
with $\phi_1=0$,  $\phi_2=0$, and  $\phi_3=\pi/2$,
and at $t_h=\Delta=1$.
The LDOS($\omega=0,j$) shows  four Majorana zero modes: (i) three single-site
MZMs at the end sites $j=1$, $31$, and $46$, while (ii) near the central region, there is one
multi-site Majorana $\chi_5$. For $V=0$, the MZM $\chi_5$ is distributed among the
central sites $j=15$, $17$, and $32$, with spectral weight $1/4$, $1/4$, and $1/2$, respectively (compared to single-site MZMs with weight 1).
 For $V=2$ the MZMs are spread over more sites compared to the $V=0$ case.
}
\label{fig5}
\end{figure}

\noindent {\bf \\ The case $ {\bf{\phi_1=0}}$, ${\bf{\phi_2=0}}$, and  ${\bf{\phi_3=\pi/2}}$\\}
  Finally, we consider the $Y$-shape Kitaev wires with phases $\phi_1=0$, $\phi_2=0$, and  $\phi_3=\pi/2$. 
This limit is also equivalent to two perpendicular Kitaev chains with phase difference of $\pi/2$ ($T$-shape wire)~\cite{Alicea,Luzie}.
The  Hamiltonian for three legs $H^{I}, H^{II}$, and $H^{III}$ takes the same form as Eqs.~5, 6, and 7, in terms of Majorana operators.
As expected, in these equations the Majorana operators $\gamma^{I}_{A,l}$, $\gamma^{II}_{B,2l+1}$, and $\gamma^{III}_{B,3l+1}$
are absent, indicating the presence of three single-site MZMs at the edge sites of the $Y$-shape Kitaev wire (see Fig.~\ref{fig5} ).
The central region, $H^{IV}$, in terms of Majorana operators becomes:
\begin{eqnarray}
H^{IV}= -\sqrt{2}i\Delta \gamma^{III}_{A,2l+2} \left( \gamma^{IV}_{A,l+1}+ \gamma^{IV}_{B,l+1}\right)\\ \nonumber
       -2i\Delta \left(\gamma^{IV}_{A,l+1} \gamma^I_{B,l} + \gamma^{II}_{A,l+2} \gamma^{IV}_{B,l+1}\right).
\end{eqnarray}
Equation 16 can be written as a $5\times 5$ matrix in the basis 
$\left[ \gamma^{II}_{A,l+2},  \gamma^{III}_{A,2l+2}, \gamma^I_{B,l}, \gamma^{IV}_{A,l+1}, \gamma^{IV}_{B,l+1}\right] $.
After diagonalizing $H^{IV}$, we obtained five eigenvalues $\left(-\sqrt{2}, \sqrt{2},-1,1,0 \right)$.
The last eigenvalue $e_5=0$ and its eigenvector $\chi_5= -\frac{1}{2} \gamma^{II}_{A,l+2} + \frac{1}{\sqrt{2}} \gamma^{III}_{A,2l+2} +\frac{1}{2}  \gamma^I_{B,l}$
has the property $\chi^{\phantom \dagger}_5=\chi_5^{\ {\dagger}}$, and $[H^{IV},\chi_5^{\phantom {IV}}]=0$,  confirming the 
presence of a  spread MZM near the junction. 
{\ {The MZM $\chi_5$ is distributed on three sites $j=15$, $17$, and $32$, showing  the multi-site nature of $\chi_5$ (see Fig.~\ref{fig5}) }}. 
The central region $H^{IV}$ can be written in a diagonal form as:
\begin{equation}
H^{IV}=2\sqrt{2}\Delta \left({\chi_2}^{{\dagger}} \chi_2 -\frac{1}{2}\right) +2 \Delta \left({\chi_4}^{{\dagger}} \chi_4 -\frac{1}{2}\right),
\end{equation}
where $ \chi_2=\frac{i}{2\sqrt{2}} \gamma^{II}_{A,l+2} + \frac{i}{2} \gamma^{III}_{A,2l+2}- \frac{i}{2\sqrt{2}}  \gamma^I_{B,l} +\frac{1}{2} \gamma^{IV}_{A,l+1} + \frac{1}{2}\gamma^{IV}_{B,l+1}$
and $\chi_4=  \frac{i}{2} \gamma^{II}_{A,l+2} +  \frac{i}{2}  \gamma^I_{B,l} -\frac{1}{2} \gamma^{IV}_{A,l+1} + \frac{1}{2}\gamma^{IV}_{B,l+1} $.
Note that the diagonalized system Hamiltonian, in case of the SC phase ($\phi_1=0$, $\phi_2=0$, and  $\phi_3=\pi/2$)
and ($\phi_1=0$, $\phi_2=0$, and  $\phi_3=0$) take a similar form (see Eqs.~15 and 17), which lead to the same energy spectrum for both cases,
 although the multi-site Majoranas wavefunctions are quite different for these two cases.
The remaining  Hamiltonians $H^{I}$,  $H^{II}$, and  $H^{III}$, after diagonalizing in terms of the ordinary $d_{k,j}$ fermionic operators,
 take similar forms as Eqs.~10, 11, and 12. In conclusion, our analytical calculations find a total of 4 MZMs. 
The three single-site MZMs are located at  edge sites in their natural positions, while a multi-site MZM $\chi_5$ is situated near the central region. 
These four MZMs results in four-fold degeneracy in the ground state of the system.

In Fig.~\ref{fig5} , we present the DMRG calculations with  $\phi_1=0$, $\phi_2=0$, and  $\phi_3=\pi/2$ for different values of $V$ using a system size $L=46$.
Similarly to the previous cases, the LDOS($\omega=0,j$) shows sharp peaks for the edge sites $j=1, 31,$ and $46$, indicating three single-site localized edge MZMs.
Interestingly, near the center the LDOS($\omega=0,j$) shows three peaks with heights $1/4$, $1/4$, and $1/2$ (compared to the edge sites) at sites $j=15, 17,$ and $32$,
respectively. These peaks in the LDOS($\omega=0,j$) indicate the presence of a multi-site 
MZM near the central site. The peak height can be explained by the special 
form of the multi-site MZM wavefunction $\chi_5= -\frac{1}{2} \gamma^{II}_{A,l+2} + \frac{1}{2} \gamma^I_{B,l}+ \frac{1}{\sqrt{2}} \gamma^{III}_{A,2l+2} $,
showing that $\chi_5$ is distributed on sites $j=15, 17,$ and $32$ with amplitudes exactly $1/4$, $1/4$, and $1/2$, respectively. Increasing the repulsion $V$,
the peak height of  LDOS($\omega=0,j$) decreases for different sites and these MZMs become exponentially localized over a few sites (Fig.~\ref{fig5}). 
 We find that the peak height of the electron and hole portions of  LDOS($\omega=0,j$) 
take the same values even for $V=2$, indicating the MZMs are quite stable against repulsive interaction for $V \le 2$. 

\noindent {\bf \\Discussion\\}
{\ {In this publication, we studied the $Y$-shaped interacting Kitaev chains using analytical and DMRG methods for different superconducting
 phases at each arm. The $Y$-shape quantum wires are potentially important to perform braiding experiments in topological quantum computations.
We  predict the unexpected presence of multi-site Majorana zero modes, in coexistence with the standard single-site Majoranas of the Kitaev model.
Based on our analytical and DMRG results: (i) For  $\phi_1=\pi$, $\phi_2=0$, and $\phi_3=0$, we predict a total of six MZMs, a large number. There are
three single-site MZMs on each edge sites, one single-site MZM at the central site, and also there are two multi-site  MZMs near the central region,
which results in three  peaks at sites $l$ (arm $I$), $l+1$ (arm $II$), and $2l+2$ (arm $III$),  with heights 2/3 (as compared to the edge sites with height 1), 
in the site-dependent LDOS calculations.    
(ii) For  $\phi_1=0$, $\phi_2=0$, and $\phi_3=0$, we find a total of four MZMs, three single-site localized at the edge sites and one multi-site MZM near the central site.
The latter is equally distributed at sites $l+2$ (arm $II$) and $2l+2$ (arm $III$), leading to two peaks with height 1/2 (compared to the edge sites)
as unveiled by the site-dependent LDOS. 
(iii) For  $\phi_1=0$, $\phi_2=0$, and $\phi_3=\pi/2$, we find a total of four MZMs as well: three 
single-site localized at the edge sites and one multi-site MZM near the center. This multi-site MZM is distributed on sites $l$ (arm $I$), $l+2$ (arm $II$), 
and $2l+2$ (arm $III$), leading to three peaks with heights $1/4$, $1/4$, and $1/2$ in the LDOS calculation.}}

{\ { Our work is motivated by
recent progress on quantum dot experiments, where the MZM modes appears at the sweet spot $t_h=\Delta$.
 Consequently, working in this idealized limit is no longer an abstract idea addressed by theory but
it has an associated reality in actual experiments. 
In this limit for the one-dimensional quantum-dots chain, the edge sites MZMs are fully localized at one site, 
to be compared to the semiconducting nanowires where $\Delta$ $\ll$ $t_h$, and the MZMs decay over many sites.
For this reason, the $Y$-shape array of quantum-dots can provide the opportunity to perform braiding experiments using only a few sites,
compared to the semiconducting nanowires that need a larger system. 
As expressed before, we found the surprising result that 
some MZMs are multi-site near the junction of the $Y$ geometry. This is {\it not} the canonical
exponential decay when
away from the sweet spot or when adding correlations, but the MZM is spread over a small
number of sites as in a ``box'' with sharp boundaries. We found the exotic result that some sites contain
1/2 of a Majorana, some 1/4 of a Majorana, and others 2/3 of a Majorana. This conclusion is in agreement with exact analytical
results at the sweet spot and confirmed numerically.}}

 {\ {The knowledge of the multi-site Majorana wave-function shape is essential 
when we exchange the MZMs near the junction in 
such quantum dot systems. Braiding requires that the Majorana wave functions do not overlap. 
Near the central region of $Y$-shape quantum wire, the junction can be also made by three dots mutually coupled to each other in a triangular geometry~\cite{Luna}
(instead of three wires coupled to one central quantum dots). Interestingly, we find that the 
 multi-site Majorana modes will still appear near the junction (with different form). Readers are referred to the SM for more detail~\cite{SM}. 
}}

 {\ {Furthermore, we compare the stability of single- and multi-site MZMs against the repulsive 
interaction $V$, by calculating the electron and hole part of $LDOS(\omega, j)$ separately.
Our DMRG results shows the single-site edge MZMs and multi-site MZMs are equally stable, 
as the peak values of LDOS$(\omega,j)$ reduce with similar rate, for moderate values of repulsive interaction.
We also checked the stability of MZMs away from the sweet spot. 
For $\Delta/t_{h}$ not too different from 1, the MZMs are quite stable and almost localized
on the same sites as in the sweet spot. For the smaller values of $\Delta$, as in semiconducting nanowires, 
the central MZMs  decays exponentially at each center over many sites and overlap with the exponentially decaying edge MZMs.}}

 {\ {We believe that our finding of multi-site MZMs, and its stability against the
 Coulomb repulsion and deviation from sweet-spot,
 will be useful to build fully functional $Y$-shape junction made 
 from array of quantum-dots~\cite{Bordin,Dvir}.
The multi-site MZMs should be observed in quantum-dots experiments, close to the sweet spots,  
using just seven quantum dots in a $Y$-shape geometry in the tunneling-conductance measurements~\cite{Dvir}. In this paper, we primarily focused on finding the physical location of the 
MZMs in a $Y$-shape geometry Kitaev chain in the $\Delta$ = $t_h$ limit realizable in quantum dots. 
In the near future, it will be 
also interesting to study MZMs in the $X$-shaped Kitaev wire,
and analyze the effect of disorder and temperature on these systems.
A recent study shows the $X$-shape wire is also quite important for the braiding process in quantum wires~\cite{Fornieri,Tong}. 
}}

\noindent {\bf \\Methods\\}
\noindent {\small \bf DMRG method\\}
In order to solve numerically the $Y$-shaped Kitaev Hamiltonian and measure observables,  
we have used the density matrix renormalization group (DMRG) method~\cite{white1992density,schollwock2005density}  with {\textsc{DMRG++}}~\cite{alvarez}.
We performed our DMRG calculations within the two-site DMRG approach, for a system size $L=46$ sites and employing $m=1500$ states, with truncation error $\le 10^{-10}$.

\noindent {\small \bf Local density-of-states\\}
We have calculated the local density-of-states $LDOS(\omega,j)$ as a function of frequency $\omega$ and site $j$, 
via the Krylov-space correction vector DMRG; for a technical review see~\cite{Nocera}.
The electron part of the LDOS$(\omega,j)$ is~\cite{Herbrych}:
\begin{equation}
LDOS^e(\omega,j)=\frac{1}{\pi} Im\left[ \left< \psi_0 \left| c_{j}^{\ {\dagger}} \frac{1}{\omega +H -(E_g-i\eta)}c_{j} \right|\psi_0 \right> \right],
\end{equation}
and the hole part of  LDOS$(\omega,j)$ is~\cite{Herbrych}:
\begin{equation}
LDOS^h(\omega)=\frac{-1}{\pi} Im\left[ \left< \psi_0 \left| c_{j} \frac{1}{\omega -H +(E_g-i\eta)}c_{j}^{\ {\dagger}} \right|\psi_0 \right> \right],
\end{equation}
where $c_{i}$ is the fermionic annihilation operator while $c^{\dagger}_{j}$ is the creation operator,
and $E_g$ is the ground state energy. We use as broadening parameter $\eta=0.1$ as in previous studies~\cite{Pandey,brad}. 
The total local density-of-states is defined as  $LDOS(\omega,j)$= $LDOS^e(\omega,j)$ +  $LDOS^h(\omega,j)$.
For the Majorana zero mode, it is expected  that the peak values of $LDOS^e(\omega,j)$ and $LDOS^h(\omega,j)$ be at or very close to $\omega=0$.

\noindent {\bf {\small \\ Data availability\\}} The data that support the findings of this study are available from the corresponding author upon request.

\noindent {\bf {\small \\ Code availability\\}} The computer codes used in this study are available at \href{https://g1257.github.io/dmrgPlusPlus/}{https://g1257.github.io/dmrgPlusPlus/.}

\noindent {\bf {\small \\Acknowledgments\\}}
The work of B.P., N.K., and E.D. was supported by the U.S. Department of Energy, Office of Science, Basic Energy Sciences,
 Materials Sciences and Engineering Division. G.~A.~ was supported by the U.S. Department of Energy,
Office of Science, National Quantum Information Science Research Centers, Quantum Science Center.

\noindent {\bf {\small \\ Author contributions\\}} B.P. and E.D. designed the project. N.K.  and B.P. carried out the analytical calculations for the $Y$-shaped Kitaev model. B.P performed the numerical DMRG calculations. G.A. developed the DMRG++ computer program. 
B.P., N.K., and E.D. wrote the manuscript. All co-authors provided useful comments and discussion on the paper.

\noindent {\bf {\small \\ Competing interests\\}} The authors declare no competing interests.

\noindent {\bf {\small \\ Additional information\\}}
Correspondence should be addressed to Bradraj Pandey ({\it bradraj.pandey@gmail.com}).



\end{document}